\documentclass[reprint, amsmath,amssymb, aps,onecolumn]{revtex4-2}

\usepackage{graphicx}
\usepackage{dcolumn}
\usepackage{bm}

\usepackage{tikz,xcolor,hyperref}
\definecolor{lime}{HTML}{A6CE39}
\DeclareRobustCommand{\orcidicon}{
	\begin{tikzpicture}
	\draw[lime, fill=lime] (0,0) 
	circle [radius=0.16] 
	node[white] {{\fontfamily{qag}\selectfont \tiny ID}};
	\draw[white, fill=white] (-0.0625,0.095) 
	circle [radius=0.07];
	\end{tikzpicture}
	\hspace{-2mm}
}
	
\foreach \x in {A, ..., Z}{\expandafter\xdef\csname orcid\x\endcsname{\noexpand\href{https://orcid.org/\csname orcidauthor\x\endcsname}
			{\noexpand\orcidicon}}
}

\begin{document}

\preprint{APS/123-QED}

\title[A unified view of elastic and elasto-inertial turbulence in channel flows at low and moderate Reynolds numbers]{A unified view of elastic and elasto-inertial turbulence in channel flows at low and moderate Reynolds numbers}

\author{Giulio Foggi Rota \orcidA{}$^1$}
\author{Christian Amor \orcidB{}$^1$}
\author{Soledad Le Clainche \orcidC{}$^2$}
\author{Marco Edoardo Rosti \orcidD{}$^1$}
\email[E-mail for correspondence: ]{marco.rosti@oist.jp}
\affiliation{$^1$ Complex Fluids and Flows Unit, Okinawa Institute of Science and Technology Graduate University, 1919-1 Tancha, Onna-son, Okinawa 904-0495, Japan.\\ $^2$ School of Aerospace Engineering, Universidad Politécnica de Madrid, E-28040 Madrid, Spain.}

\begin{abstract}
The chaotic flow of elastic fluids at low Reynolds number ($Re$) is typically distinguished into elasto-inertial and elastic turbulence (EIT/ET). However, the clear separation among these two turbulent regimes in parallel flows with a gradual $Re$ decrease remains elusive. Here, spanning $Re$ over four orders of magnitude, we investigate the statistical and structural nature of the flows computed by fully resolved 3D numerical simulations, revealing that as inertia becomes sub-dominant all flows share similar dynamical properties. Supported by the results, we thus propose a unified view of fully developed ET and EIT in parallel flows.
\end{abstract}

\maketitle

%Introduction

\begin{figure}
%\begin{tikzpicture}
%	\node[frame_2800, inner sep=0pt] (Re2800) at (0,7.5) {\includegraphics[width=.44\textwidth]{view_Re2800.png}};
%	\node[text=color2800,align=left,rotate=270] at (4.3,7.5) {$2800$};
%	\node[frame_1500, inner sep=0pt] (Re1500) at (0,6) {\includegraphics[width=.44\textwidth]{view_Re1500.png}};
%	\node[text=color1500,align=left,rotate=270] at (4.3,6) {$1500$};
%	\node[frame_500, inner sep=0pt] (Re500) at (0,4.5) {\includegraphics[width=.44\textwidth]{view_Re500.png}};
%	\node[text=color500,align=left,rotate=270] at (4.3,4.5) {$500$};
%	\node[frame_50, inner sep=0pt] (Re50) at (0,3) {\includegraphics[width=.44\textwidth]{view_Re50.png}};
%	\node[text=color50,align=left,rotate=270] at (4.3,3) {$50$};
%	\node[frame_5, inner sep=0pt] (Re5) at (0,1.5) {\includegraphics[width=.44\textwidth]{view_Re5.png}};
%	\node[text=color5,align=left,rotate=270] at (4.3,1.5) {$5$};
%	\node[frame_05, inner sep=0pt] (Re05) at (0,0) {\includegraphics[width=.44\textwidth]{view_Re05.png}};
%	\node[text=color05,align=left,rotate=270] at (4.3,0) {$0.5$};
	
%	\node[inner sep=0pt] (Top) at (4.6,7.5) {};
%	\node[inner sep=0pt] (Bottom) at (4.6,0) {};
%	\draw[->,thick] (Top) -- (Bottom) node [fill=white, pos=0.5, above] {$Re$};
%\end{tikzpicture}
\includegraphics[width=.58\textwidth]{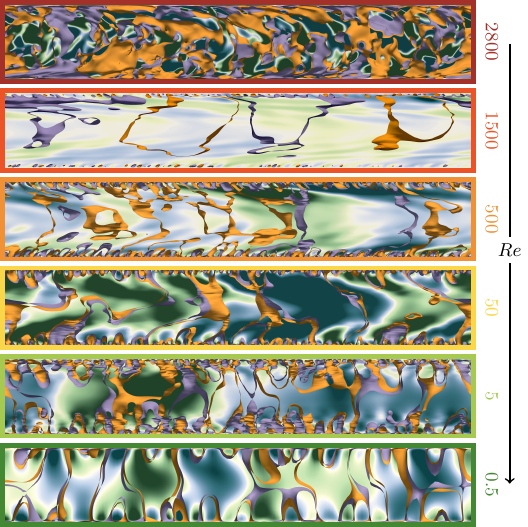}
\caption{\textbf{Side views} of the channel, with the mean flow going from left to right. We show the spanwise velocity fluctuations normalized with the mean turbulent kinetic energy, $w'/\sqrt{K}$, ranging from -0.3 (blue) to 0.3 (green); superimposed are the local isosurfaces of the wall-normal fluctuations, $v'/\sqrt{K}$, at $-0.05$  (orange) and $+0.05$ (violet).}
\label{fig:cover}
\end{figure}

Turbulence is an open problem in fluid transport applications, as it significantly enhances the drag and energy expenditures. Nevertheless, it is beneficial in mixing problems since it introduces an additional diffusion mechanism on top of the classical viscous one. Indeed, viscous diffusion is often too slow for practical applications, but it is the only mechanism operating in low Reynolds number flows, which are typically laminar. The Reynolds number, $Re$, is the non dimensional parameter conventionally adopted to describe fluid flow, representing the relative importance of inertia compared to viscosity. Micro-scale mixing in low Reynolds number flows is a widely investigated research topic \cite{stroock-etal-2002,ober-foresti-lewis-2015,han-lafiandra-devoe-2022} for its high biological relevance, e.g., in drug delivery applications \cite{vargason-anselmo-mitragotri-2021} and in the analysis of specimens for disease diagnosis \cite{yager-etal-2006}. Among the various mixing techniques, one is to exploit elastic turbulence \cite{groisman-steinberg-2001-2}, a peculiar chaotic state unique of viscoelastic fluids which are characterised by the coexistence of viscous and elastic effects.

The description of viscoelastic fluid flows necessitates an additional parameter accounting for the ratio of the elastic and flow time scales: the Deborah number, $De$. Comparing the Deborah to the Reynolds number allows to identify distinct regimes of motion: when inertial effects are dominant turbulent drag reduction \cite{serafini-etal-2022} (TDR) occurs, while elastic turbulence \cite{groisman-steinberg-2001-1} (ET) is observed for dominant elasticity. In between these two extrema, an intermediate state dubbed elasto-inertial turbulence \cite{doubief-terrapon-hof-2023} (EIT) has been usually identified, where both elastic and inertial effects are relevant.
There is no universal consensus on the definitions of EIT and ET; yet, for the purposes of our investigation, we follow \citet{groisman-steinberg-2001-1} and denote with ET the chaotic flow observed in polymer solutions at small values of $Re$ and large values of $De$ ($Re\lesssim De$). 
Furthermore, according to the definition given by \citet{samanta-etal-2013} recently adopted also by \citet{doubief-terrapon-hof-2023}, we dub EIT the chaotic motion controlled by the elastic stresses that can set in at lower Reynolds number than inertial turbulence, in a parameter range where inertia cannot be neglected ($Re\approx De$).
Similarly, \citet{page-dubief-kerswell-2020} defined EIT as a new turbulent flow state observed at modest inertia and elasticity, while \citet{khalid-shankar-subramanian-2021} dubbed EIT the ensuing flow state to which pipe and channel flows of sufficiently elastic polymer solutions transition from the laminar state at Reynolds numbers substantially lower than the typical Newtonian threshold.

EIT occurs at significantly lower Reynolds numbers than TDR, where the laminar flow of a Newtonian fluid would be linearly stable, and it is characterised by different dynamical properties \cite{samanta-etal-2013}. When inertial effects vanish, the flow can still become turbulent (ET) due to the onset of different elastic instabilities. 
It is in fact well established that different instability mechanisms are responsible for the transition to a chaotic state at moderate \cite{garg-etal-2018-1} and vanishing \cite{berti-boffetta-2010} values of $Re$, thus motivating the distinction between EIT and ET in the stability framework \cite{doubief-terrapon-hof-2023}. 
In curvilinear wall bounded flows, ET is triggered and sustained by a force directed as the flow curvature, called hoop stress \cite{bird-1987, groisman-steinberg-2001-1}. Recent experiments \cite{jha-steinberg-2021, li-steinberg-2023} and numerical simulations \cite{song-etal-2022, lellep-linkmann-morozov-2024} have highlighted that ET can also be attained in parallel shear flows, but a comprehensive characterisation of its mechanisms is still elusive. After the discovery of a linear elastic instability in the parallel Kolmogorov flow \cite{boffetta-etal-2005}, the weakly non-linear regime following the instability has been studied \cite{bistagnino-etal-2007} and observed to culminate in ET \cite{berti-etal-2008}. Arrowhead-shaped traveling elastic waves were first observed in this scenario \cite{berti-boffetta-2010} and later found also in the bulk of channel flows in the EIT \cite{sid-terrapon-dubief-2018, garg-etal-2018-1, dubief-etal-2022, buza-etal-2022} and ET regimes \cite{morozov-2022}, suggesting a possible continuous path between the two \cite{doubief-terrapon-hof-2023} and hinting towards the existence of a unique sustainment mechanism common to EIT and ET \cite{khalid-shankar-subramanian-2021}.
Channel flows at moderate to vanishing inertia have been recently observed to share also a different instability \cite{beneitez-page-kerswell-2023}, localised close to the walls. 
Nevertheless, due to a lack of measurements, it is still unclear if the fully developed three-dimensional EIT and ET chaotic states are the same, connected or completely decoupled. 

In this Letter we show that \textit{fully developed EIT and ET in planar channel flows below $Re\approx2800$ share a similar structure}, notwithstanding the instabilities leading to them. 
Addressing the fundamental question \textit{``Is EIT simply an inertial version of ET or does inertia play a more fundamental role?"} posed by \citet{doubief-terrapon-hof-2023}, we thus clarify the relation between EIT and ET within this setup.
Our fully-resolved three-dimensional numerical simulations, spanning Reynolds numbers over four orders of magnitude (as in FIG.~\ref{fig:cover}), allow us to bridge the different regimes (TDR, EIT and ET) and clarify their relation. For every Reynolds number considered in this study we observe the system to converge towards a self-sustained chaotic state even when the flow of a Newtonian fluid would instead be laminar, and we identify only two different regimes, one at large Reynolds number (TDR) and one at smaller ones (ET). 

%Methods

To show this, we simulate the flow of a viscoelastic incompressible fluid between two indefinite planar walls at a distance of $2h$ from each other, described in a Cartesian coordinate frame with the $x$, $y$ and $z$ axes aligned with the streamwise, wall-normal and spanwise directions, respectively. Variables and symbols are listed in section V of the Supplemental Material.
The system is governed by the conservation of mass and momentum,
\begin{equation}
	\mathbf{\nabla} \cdot \mathbf{u}=0,
\label{eq:incompressibility}
\end{equation}
\vspace{-.5cm}
\begin{equation}
	\rho \left( \partial_t \mathbf{u}+\mathbf{\nabla} \cdot \mathbf{u} \mathbf{u} \right)=-\mathbf{\nabla}p
	+\mu_s\mathbf{\nabla}\cdot(\mathbf{\nabla u}+(\mathbf{\nabla u})^T)+\rho \mathbf{f} + \mathbf{\nabla}\cdot\mathbf{T}, 
\label{eq:momentum}
\end{equation}
where $\mathbf{u}$ is the velocity, $\rho$ the density, $p$ the pressure and $\mu_s$ is the dynamic viscosity of the Newtonian solvent in which the polymer molecules are dissolved. $\mathbf{f}$ represents the forcing term needed to drive the flow at a constant flow rate, while $\mathbf{T}$ is the tensor accounting for the non-Newtonian behaviour. In our study, $\mathbf{T}$ derives from a finitely extensible nonlinear elastic model with Peterlin's closure (FENE-P \cite{peterlin-1966}, equivalent to a bead-spring-type model \cite{vincenzi-etal-2007}) and relates to the polymer conformation tensor, $\mathbf{C}$, as $ \mathbf{T}(\mathbf{C})= (\mu_p/\lambda)(f(tr(\mathbf{C}))\mathbf{C}-(L_{max}^2\mathbf{I})/(L_{max}^2-3))$, where $L_{max}$ denotes the maximum extensibility of the polymer chains, $\mathbf{I}$ is the tensorial identity, $\mu_p$ is the dynamic viscosity of the polymer, and $\lambda$ is its relaxation time. In particular, $f(s)=L_{max}^2/(L_{max}^2-s)$. Finally, $\mathbf{C}$ evolves in time according to its own transport equation
\begin{equation}
	\partial_t \mathbf{C}+(\mathbf{u}\cdot \mathbf{\nabla})\mathbf{C} - \mathbf{C}\cdot\mathbf{\nabla u} - (\mathbf{\nabla u})^T\cdot \mathbf{C} = -\displaystyle\frac{1}{\lambda}
	(f(tr(\mathbf{C}))\mathbf{C}-\mathbf{I}).
\label{eq:conformation}
\end{equation}

The regime of motion of the viscoelastic fluid is identified by the specification of the Reynolds and the Deborah numbers, along with the solvent-to-total viscosity ratio, $\beta$, and $L_{max}$. In this work we adopt the definition of ``bulk'' Reynolds number, $Re=h U/\nu$, based on the mean velocity within the channel, $U$, and the total kinematic viscosity of the polymeric solution, $\nu=(\mu_s+\mu_p)/\rho$; we therefore decrease $Re$ in $\{ 2800, 1500, 500, 50, 5, 0.5\}$ starting from the upper-bound for EIT reported by \citet{doubief-terrapon-hof-2023}. Furthermore, we complement our results with those available in the literature by \citet{lellep-linkmann-morozov-2024} at even smaller Reynolds number, $Re=0.01$.
The Deborah number is defined as $De=\lambda U/h$ and imposed equal to $50$ throughout the whole study, while  $L_{max}$ and $\beta=\mu_s/(\mu_s+\mu_p)$ are set to $500$ and $0.9$, respectively. 
We start each simulation from a fully developed field of the closest case at a higher Reynolds number, and advance it in time according to the methods detailed in section I of the Supplemental Material \textit{(see the Supplemental Material for technical details on the numerical methods. There, we also provide further insight on the flow statistics, elucidate the calibration of the modal decomposition algorithm and investigate the dependence of the results from the polymer concentration. Relevant references are contained therein \cite{doubief-terrapon-hof-2023,abdelgawad-cannon-rosti-2023,kim-moin-1985,fattal-kupferman-2005,devita-etal-2018,sugiyama-etal-2011,beneitez-etal-2024,deangelis-casciola-piva-2002,lellep-linkmann-morozov-2024,pope-2000,serafini-etal-2022,vega-leclainche-2020,schmid-2010,dubief-etal-2022,couchman-etal-2024,buza-page-kerswell-2022,beneitez-page-kerswell-2023})}.  
Our parameters are chosen consistently with previous stability analysis \cite{buza-etal-2022} and yield values of the elasticity number $El = De / Re$ in $\{0.02,0.03,0.1,1,10,100\}$, denoting how the investigated range equally encompasses cases dominated by inertia ($El \ll 1$) and elasticity ($El \gg 1$), as well as the intermediate range.
The results presented in the following have been replicated also with the Oldroyd-B liquid, thus proving their robustness against the polymer model adopted; furthermore, we have tested the effect of different viscosity ratios $\beta$, as detailed in section IV of the Supplemental Material.
  
%Results

\begin{figure}
\includegraphics[width=\textwidth]{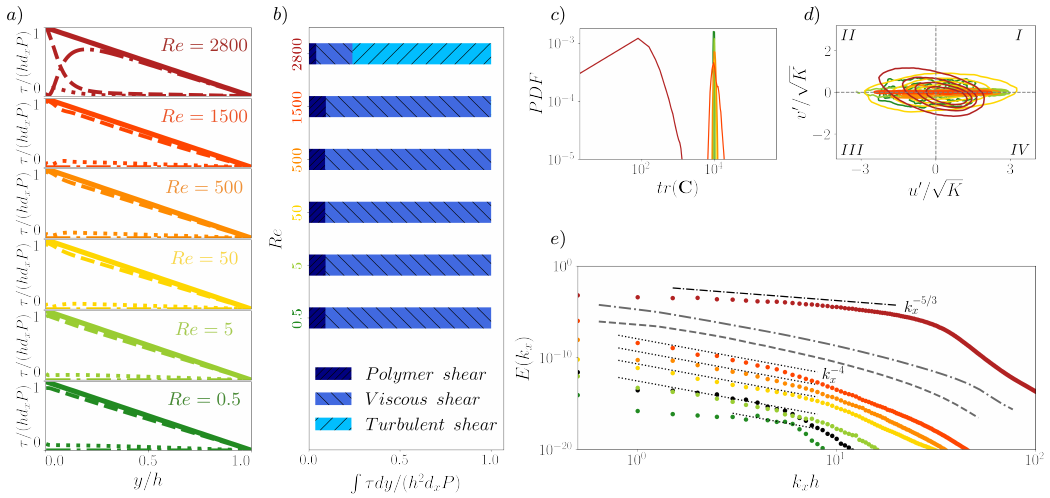}
\caption{\textbf{Flow statistics.} In $(\mathbf{a})$ we report the viscous (dashed), turbulent (dashed-dotted) and polymer (dotted) shears, summing up to the total shear stress $\tau$ (solid line). At $Re \leq 1500$ viscosity and elasticity dominate throughout the whole half-channel height $h$.
All the contributions integrated along the wall-normal direction, in ($\mathbf{b}$), are thus dominated by the viscous and polymer shears, but for the case at $Re=2800$, dominated by the turbulent shear. 
Consistently, the polymer stretching is enhanced, as shown in $(\mathbf{c})$ by the PDFs of the trace of $\mathbf{C}$ at the channel centerline: their barycenters are far from $tr(\mathbf{C})=3$.
$(\mathbf{d})$ shows the J-PDFs of the normalized streamwise and wall-normal velocity fluctuations at $y=h/2$, at $\{0.1,0.3,0.5,0.7\}$ of their maximum. The typical turbulent shape is recovered at $Re=2800$, while in all the other cases the fluctuations decorrelate. 
We show the TKE spectra $(\mathbf{e})$ with an arbitrary vertical shift, to prevent their overlapping, along with Oldroyd-B data at $Re=5$ (black circles) and data from \citet{lellep-linkmann-morozov-2024} at $Re=0.01$ and $De=80$/$150$ as a gray dashed/dash-dotted line, respectively.} The $-4$ slope contrasts sharply with the $-5/3$ decay recovered in TDR, pointing towards a unified view of EIT and ET.
\label{fig:statistics}
\end{figure}

At all the Reynolds numbers but the highest one, qualitatively similar coherent fluctuations of the velocity are superimposed to a mean flow which only slightly departs from the laminar one (more details on this in section II of the Supplemental Material). As visible in FIG.~\ref{fig:cover}, large structures span the center of the domain, while smaller ones characterize the region near the wall. All of them exhibit a quasi-periodic spatial arrangement superimposed to random perturbations. A more quantitative characterization of the flow fields is attained looking at the balance of the total shear stress, $\tau$, along the half-channel height, $h$. The conservation of linear momentum requires the wall-normal derivative of $\tau$ to balance the driving pressure gradient, $d_y\tau=d_xP$, imposing a linear trend of $\tau$ with $y$. Three different contributions sum together in the total shear stress: a polymeric one due to the shearing component of the viscoelastic stresses $\mathbf{T}$ ($\langle T_{xy} \rangle$, where angle brackets denote averaging in time and along the homogeneous directions), a purely viscous one coming from the mean velocity profile $\langle u \rangle$ ($\rho \nu \partial_y \langle u \rangle$), and a turbulent one caused by the correlated fluctuations of the streamwise and wall-normal velocities ($-\rho  \langle u'v' \rangle$, where the primes indicate fluctuations). These contributions are reported in the plots of FIG.~\ref{fig:statistics}$a$.

As expected for the TDR regime at $Re=2800$, the viscoelastic contribution remains small almost everywhere and viscosity is mostly relevant close to the wall, while the turbulent contribution dominates the center of the channel. For $Re \leq 1500$, instead, the shear stress balance is completely different, as the polymer and viscous shears become relevant throughout the channel and the turbulent one is depleted. To better weight the relative importance of the different contributions of the shear stress balance, we integrate the momentum balance along the wall-normal direction ($\int{(\langle T_{xy} \rangle+\rho \nu \partial_y \langle u \rangle -\rho  \langle u'v' \rangle)dy}=h^2 \enspace d_xP$), as represented graphically in FIG.~\ref{fig:statistics}$b$. The turbulent shear is the most relevant component at $Re=2800$, while it is practically zero in all the other cases; those are dominated by the viscous shear, along with a non-negligible polymeric contribution (individually shown in section II of the Supplemental Material). Significantly, the relative amplitude of the different contributions remains unchanged for all the Reynolds numbers (ET) but the highest one (TDR).

The previous results warrant two important clarifications. First, the relative small contribution of the polymers does not mean that the polymers are not stretched. Indeed, the polymer molecules are significantly stretched regardless of the case considered, as shown by the probability density functions (PDF) of the trace of $\mathbf{C}$ reported in FIG.~\ref{fig:statistics}$c$. Also here, a clear separation can be appreciated between the case at $Re=2800$ and all the others. Second, the vanishing magnitude of the turbulent shear when $Re\leq 1500$ does not mean that the turbulent fluctuations disappear, but rather suggests that the streamwise and wall-normal velocity fluctuations decorrelate. We therefore compute the joint probability density function (J-PDF) of the streamwise and wall-normal velocity fluctuations at a wall distance of $h/2$, reported in FIG.~\ref{fig:statistics}$d$, where velocities are made dimensionless with the mean turbulent kinetic energy in the channel, $K$, for a better comparison of the different Reynolds numbers. 
In the high Reynolds case the isolines exhibit an oval shape with the major axis inclined downward, denoting a negative correlation among the two velocity components, and ejections ($u'<0$, $v'>0$) dominate over sweeps ($u'>0$, $v'<0$), analogously to the Newtonian case \cite{tardu-2014}.
At this wall distance, the J-PDF peaks in the fourth quadrant.
This picture completely changes when decreasing the Reynolds number, as the streamwise velocity decorrelates from the wall-normal one and its fluctuations heavily dominate in intensity; the J-PDF consequently appears flattened along the x-axis, explaining the apparent disappearance of the turbulent shear stress from FIG.~\ref{fig:statistics}.

The features of the low Reynolds flow fields described up to this point would appear compatible with an incomplete relaminarisation of the flow; nevertheless, its turbulent nature is confirmed by the full multi-scale Fourier's spectra $E$ of the turbulent kinetic energy (TKE) at the centerline of the channel. Those are characterised by a power-law decay of $E(k_x) \sim k_x^{-4}$ over roughly a decade of streamwise wave numbers $k_x$ (FIG.~\ref{fig:statistics}$e$), in good agreement with recent observations in a similar setup at lower Reynolds number \cite{lellep-linkmann-morozov-2024} and in homogeneous isotropic elastic turbulence \cite{singh-etal-2024}. 
The wavenumber range in which the scaling is observed extends moving from $Re=1500$ to $Re=50$, yet it appears significantly shrank at $Re=0.5$, suggesting a tendency of the flow to re-laminarise at even lower Reynolds numbers. This is consistent with recent stability analysis results \cite{buza-page-kerswell-2022,buza-etal-2022} considering our same polymer model (FENE-P) and same parameter range (Reynolds number, Deborah number, polymer concentration and maximum extensibility).
In the TDR regime at $Re=2800$, instead, our data exhibit a tendency towards the conventional $E(k_x) \sim k_x^{-5/3}$ trend expected for Newtonian turbulence at high Reynolds number. Once more, no significant difference is observed among all the low Reynolds number cases; the robustness of this statement to significant variations of the polymer concentration is detailed in section IV of the Supplemental Material. Our results therefore point towards a unified view of EIT and ET in three-dimensional fully developed planar channel flows below $Re\approx2800$.

\begin{figure}
%\begin{tikzpicture}
	
%	\node[inner sep=0pt] (spectra) at (0,0) {\includegraphics[width=.47\textwidth]{DMDv4.png}};
%	\node[align=left] at (-4,2) {$a)$};
%	\node[align=left] at (-1.4,2) {$b)$};
%	\node[align=left] at (-1.4,0) {$c)$};
	
%\end{tikzpicture}
\includegraphics[width=.6\textwidth]{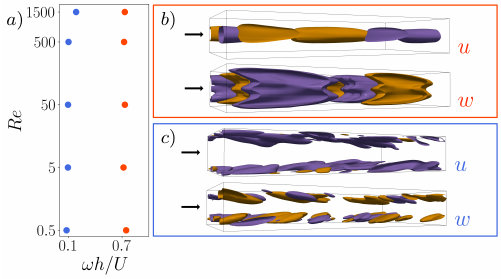}
\caption{\textbf{Modal decomposition}, showing the temporal frequencies $(\mathbf{a})$, $\omega$, of the two most dynamically relevant modes $(b,c)$ for five different values of the Reynolds number. The uncertainty of the decomposition algorithm calibration process lays within the symbols (see section III of the Supplemental Material). The colored panels~$\mathbf{b,c}$ report the typical shapes of the streamwise, $u$, and spanwise, $w$, velocity components of the modes (sampled from the case at $Re=50$), where positive isosurfaces are colored in purple and negative isosurfaces in orange. The identification of two dominant modes common to all the low Reynolds number cases (consistent with previous stability analysis results) corroborates the unified view of EIT and ET.} 
\label{fig:dmd}
\end{figure}

To further confirm this scenario, we apply a purely data-driven algorithm to identify and extract the coherent structures in the flow at different Reynolds numbers. In particular, we adopt the higher order dynamic mode decomposition \cite{vega-leclainche-2020}, a dimensionality reduction algorithm which provides a finite set of modes, each oscillating in time with a specific frequency $\omega$. Past stability analysis of viscoelastic two-dimensional channel flows suggests that travelling waves originated at a relatively high-$Re$ bifurcation \cite{garg-etal-2018-1} persist in EIT \cite{page-dubief-kerswell-2020,khalid-etal-2021,dubief-etal-2022,buza-etal-2022} and ET \cite{buza-page-kerswell-2022,morozov-2022}. These structures are mainly located away from the walls, in the central region of the channel, and are sometimes termed ``arrowheads" \cite{page-dubief-kerswell-2020}. In ET, these coexist with a recently discovered wall-localised linear instability caused by the presence of small but non-zero diffusivity of the polymer stresses \cite{beneitez-page-kerswell-2023,couchman-etal-2024}. The diffusive instability perturbs the flow close to the walls, in the region where the travelling waves convert most of the kinetic energy into elastic energy \cite{beneitez-etal-2024}. Our analysis suggests that all the turbulent flows at $Re \leq 1500$ can be described by the superposition of two main sets of modes for all the Reynolds numbers but the highest one, once more corroborating the unified view of EIT and ET suggested by the analysis above. The frequencies of these modes are shown in FIG.~\ref{fig:dmd}$a$ as a function of the Reynolds number, and their typical shapes are reported in FIG.~\ref{fig:dmd}$b$ and $c$. We observe that the fast modes are mainly located in the center of the channel, while the slow modes are localised close to the walls; their temporal frequencies do not undergo a significant variation with the Reynolds number. 
The modes we identify are thus unrelated to the turbulent structures observed by \citet{choueiri-etal-2021} in EIT, which exhibit a marked dependence on the Reynolds number and persist up to $Re=10000$, yet their shapes and locations are reminiscent of the results from stability analysis \cite{buza-page-kerswell-2022,beneitez-page-kerswell-2023} and recent simulations \cite{lellep-linkmann-morozov-2024}. 

%Conclusions

In this work we have simulated the turbulent flow of a viscoelastic fluid in a planar channel setup across a wide range of Reynolds numbers from TDR to ET, reporting instabilities and flow structures consistent with the available literature \cite{page-dubief-kerswell-2020,beneitez-page-kerswell-2023,doubief-terrapon-hof-2023}.  
Our results identify two distinct turbulent regimes. 
First, one at large Reynolds number dominated by the turbulent shear stresses with the classical energy spectra decay predicted by Kolmogorov's theory \cite{kolmogorov-1941} $E\sim k_x^{-5/3}$, which corresponds to the extensively investigated TDR regime \cite{lumley-1969}.
Next, one at lower Reynolds number with highly elongated polymer chains producing strong polymer stresses with the energy spectra showing a fast fall-off $E\sim k_x^{-4}$, consistently with recent results in ET even at Reynolds numbers lower than those here \cite{singh-etal-2024,lellep-linkmann-morozov-2024}.
We do not find any other fully-developed state in between these two, and conclude that the peculiar chaotic state attained in viscoelastic planar channel flows remains unchanged by a variation in the Reynolds number once inertial effects are sub-dominant. Thus, within this set-up, fully-developed EIT \cite{doubief-terrapon-hof-2023} below $Re\approx2800$ and ET \cite{steinberg-2021} share a similar structure.

\begin{acknowledgments}
The research was supported by the Okinawa Institute of Science and Technology Graduate University (OIST) with subsidy funding to M.E.R. from the Cabinet Office, Government of Japan. Part of this work was done during the 2023 Madrid Turbulence Workshop, organised by Prof. J. Jiménez and made possible by ERC, Caust grant \textit{ERC-AdG-101018287}. M.E.R. acknowledges funding from the Japan Society for the Promotion of Science (JSPS), grants 24K17210 and 24K00810. All authors acknowledge the computational resources provided by the Scientific Computing section of the Research Support Division at OIST and by RIKEN, under the HPCI System Research Project grants \textit{hp220099} and \textit{hp210269}. M.E.R. and G.F.R. acknowledge the insightful discussion with Prof. R. R. Kerswell and Dr. M. Beneitez at the University of Cambridge.
\end{acknowledgments}

\pagebreak
%\bibliography{./Wallturb.bib}

\end{document}

\preprint{APS/123-QED}

\begin{center}
	\textit{Supplemental Material for}
\end{center}

\title{A unified view of elastic and elasto-inertial turbulence in channel flows at low and moderate Reynolds numbers}

\author{Giulio Foggi Rota \orcidA{}$^1$}
\author{Christian Amor \orcidB{}$^1$}
\author{Soledad Le Clainche \orcidC{}$^2$}
\author{Marco Edoardo Rosti \orcidD{}$^1$}
\email[E-mail for correspondence: ]{marco.rosti@oist.jp}
\affiliation{$^1$ Complex Fluids and Flows Unit, Okinawa Institute of Science and Technology Graduate University, 1919-1 Tancha, Onna-son, Okinawa 904-0495, Japan.\\ $^2$ School of Aerospace Engineering, Universidad Politécnica de Madrid, E-28040 Madrid, Spain.}

\maketitle

\setcounter{table}{0}
\makeatletter 
\renewcommand{\thetable}{S\@arabic\c@table}
\makeatother

\setcounter{figure}{0}
\makeatletter 
\renewcommand{\thefigure}{S\@arabic\c@figure}
\makeatother

\setcounter{equation}{0}
\makeatletter 
\renewcommand{\theequation}{S\@arabic\c@equation}
\makeatother

\section{Direct numerical simulation}

The direct numerical simulation (DNS) of a viscoelastic fluid flow in the ET regime poses significant challenges.
Both the high accuracy required to the numerical schemes adopted for a correct integration of the governing equations and the computational power needed to ensure an adequate resolution make this kind of simulations a remarkable endeavour \cite{doubief-terrapon-hof-2023}.  
In this study, the problem is accurately and efficiently tackled by means of our well validated solver \textit{Fujin} (\url{https://www.oist.jp/research/research-units/cffu/fujin}, \cite{abdelgawad-cannon-rosti-2023}).
We adopt a staggered uniform Cartesian grid and discretize the governing equations in space according to a second-order central finite-difference scheme; a second-order Adams–Bashforth scheme is chosen for time integration, coupled with a fractional step method \cite{kim-moin-1985}. 
At every time step, we find a pressure field satisfying the divergence-free constraint with an efficient spectral Poisson solver, while the overall parallelisation of the code relies on the \textit{2decomp} library along with the message passing interface (MPI) protocol.
The transport equation for the polymer conformation tensor $\mathbf{C}$ requires additional care: we tackle it in a logarithmic formulation \cite{fattal-kupferman-2005,devita-etal-2018} to overcome the notorious high Deborah numerical instability and resort to a high-order weighted essentially non-oscillatory (WENO) scheme \cite{sugiyama-etal-2011} to treat the upper-convected derivative constituting the left-hand side.
In this way we are able to avoid the explicit introduction of a stress-diffusion term, which therefore does not require us to specify any boundary condition \cite{beneitez-etal-2024}.
Instead, for the fluid, no-slip and no-penetration boundary conditions are imposed at the upper and lower walls, while periodicity is enforced along the homogeneous $x$ and $z$ directions.
In the production simulations, we adopt a computational domain of size $4\pi h \times 2h \times 2\pi h$ in the $x$, $y$ and $z$ directions, respectively, across which we uniformly distribute $512 \times 1024 \times 256$ grid points. 
For the case at $Re=5$, nevertheless, we reduce our resolution to $256 \times 512 \times 128$ points and further decrease to $384$ points along $y$ at $Re=0.5$ to accelerate the otherwise prohibitively slow convergence to a fully developed turbulent state.
The resolution of our high Reynolds simulations is larger than that employed in previous numerical studies \cite{deangelis-casciola-piva-2002}, while the adequacy of a comparable grid in the lower Reynolds cases has already been tested \cite{lellep-linkmann-morozov-2024}.

\pagebreak

\section{A comparison of Newtonian turbulence, TDR and ET}

Highlighting the similar structure of fully developed ET and EIT in channel flows, our Letter leaves room for three different turbulent states: ET, TDR and Newtonian turbulence (NT).
To better investigate their relation, we have performed a fully-resolved simulation of the turbulent channel flow of a Newtonian fluid at $Re=2800$, resorting to the same numerical setup adopted in the TDR case. This further analysis allows us to justify the drag reduction claim on which TDR lays its foundations, and offers us the opportunity to clarify how ET stands away from it.

\begin{figure}[H]
\centering
\includegraphics[scale=0.46]{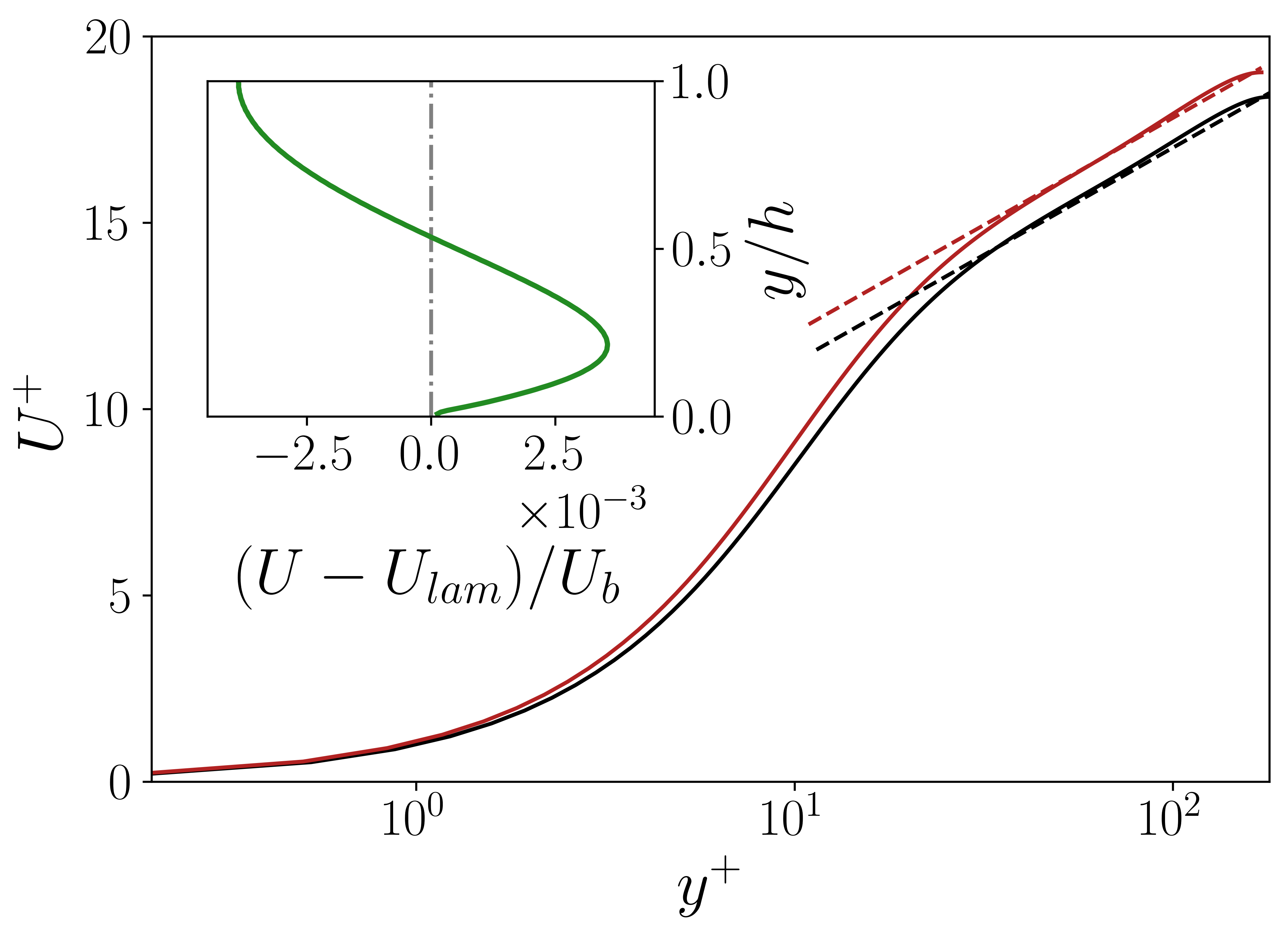}
\includegraphics[scale=0.455]{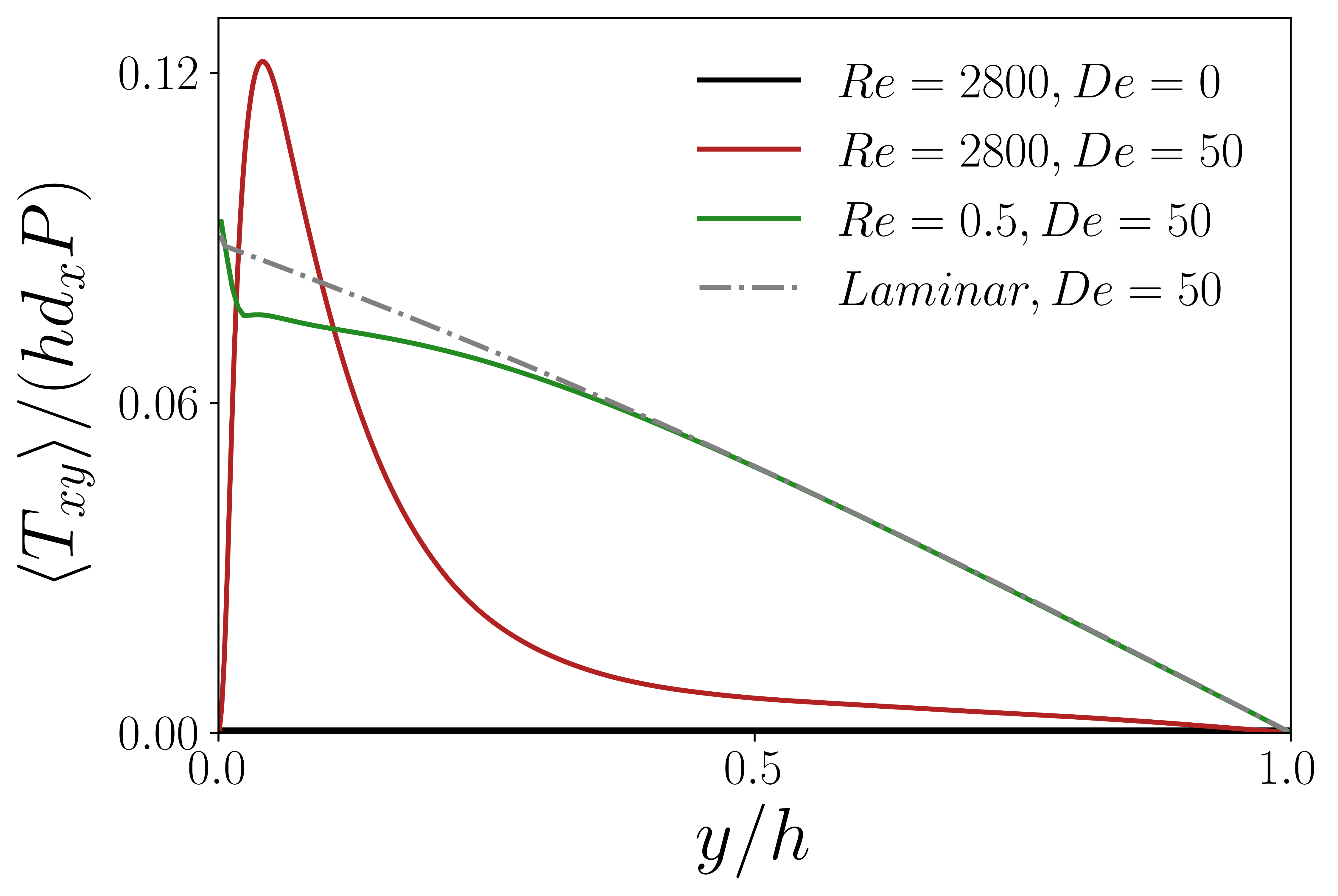}
\put(-480,155){a)}
\put(-240,155){b)}
\caption{\textbf{Mean flow properties in ET, TDR and NT.} In panel~$\mathbf{a}$ we report the mean velocity profiles scaled in wall units for NT and TDR. We fit both curves with the log-law $u^+=\frac{1}{\kappa}\log{y^+}+B$ (dashed curves), setting $\kappa=0.4$. Drag reduction is confirmed as the TDR profile appears shifted above the NT one. We also show, in the inset, the deviation between the mean velocity profile found in ET at $Re=0.5$ and the laminar one: the average streamwise velocity is depleted in the bulk of the channel and increased towards the wall, compared to the laminar case.
Panel~$\mathbf{b}$ reports the trends of the viscoelastic shear stress throughout the channel. In TDR it peaks where the turbulent activity is maximum and decays towards the center of the channel, while in ET it approaches the laminar trend with a significant deviation next to the wall. The shape of the velocity profile in TDR is akin to that in NT, but for the manifestation of drag reduction in the offset between the two. ET, instead, stands aside from the other turbulent states both in terms of mean velocity profile and viscoelastic shear stress.}
\label{fig:fProp}
\end{figure}

 We first look at the mean profiles of the streamwise velocity, reported in panel~$a$ of FIG.~\ref{fig:fProp} and scaled in wall units \cite{pope-2000}, denoted with a plus superscript. 
A logarithmic range emerges for both the NT ($Re=2800, De=0$) and TDR ($Re=2800, De=50$) cases, with the vertical shift between the two indicating that the latter is characterised by less drag. We in fact confirm a decrease in the skin-friction coefficient of $\sim10\%$ \cite{serafini-etal-2022}. 
Reporting the profile attained in ET ($Re=0.5, De=50$) on the same plot of the others does not prove informative, as it deviates significantly from the plug-like shape observed there and it instead approaches the laminar parabola.
We thus show in the inset of FIG.~\ref{fig:fProp}$a$ the deviation between the mean velocity profile found in ET and the laminar one.
The difference is moderate but finite (as also shown in FIG.~1F of \citet{lellep-linkmann-morozov-2024}), and quantifies the effect of the elastic fluctuations on the bulk flow. 
In particular, the average streamwise velocity is depleted above $y/h \approx 0.5$ and increased below that. 

The macroscopic contrast between the flow profiles observed in TDR  and ET relates to a macroscopic difference in the action exerted by the polymer on the fluid, quantified by the viscoelastic shear stress $\langle T_{xy} \rangle$.
That is shown, along with the other shearing contributions, in FIG.~2 of the Letter, yet its modest magnitude prevents the comparison of its trend across the channel between different cases. 
We thus show $\langle T_{xy} \rangle$ alone in panel~$b$ of FIG.~\ref{fig:fProp}.
While in NT it is clearly nullified, in TDR it peaks close to the location of maximum turbulent activity and decays to a small value in the bulk of the flow. 
In ET, instead, $\langle T_{xy} \rangle$ is significant throughout the channel and remains close to the laminar curve above $y/h \approx 0.5$, with an appreciable deviation below that.
Such trend appears consistent with the increase of the velocity in the region near the wall, observed in the inset of FIG.~\ref{fig:fProp}$a$.

Comparing the mean velocity profiles in NT, TDR and ET we have assessed their similarity across NT and TDR, further verifying the drag reduction claim. ET, instead, is not akin to any of the other turbulent states in terms of mean flow and viscoelastic stresses, and constitutes a unique and peculiar chaotic flow condition.

\pagebreak

\section{Calibration of the modal decomposition algorithm}
\begin{figure}[H]
\centering
\includegraphics[scale=0.48]{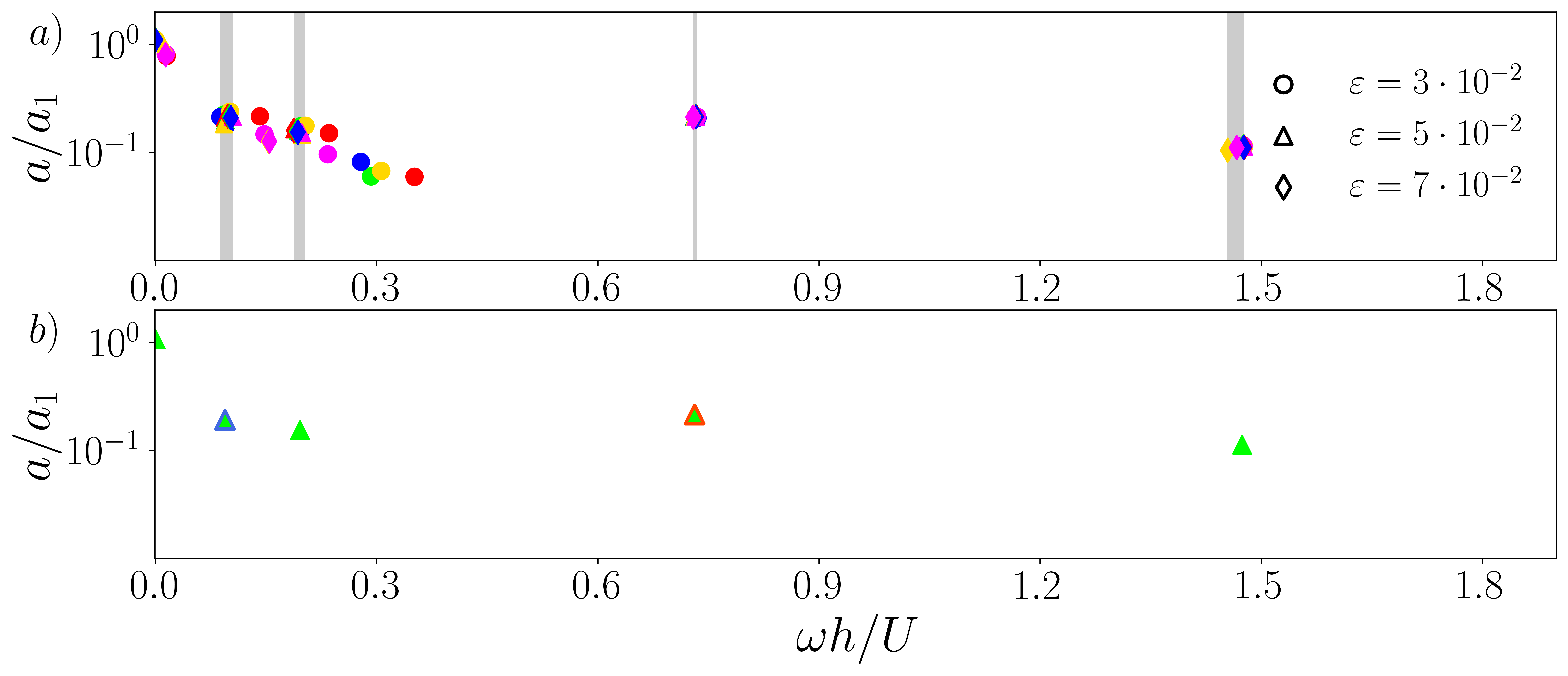}
\caption{ \textbf{HODMD calibration.} We show the calibration process for the case at $Re = 50$. In panel~$\mathbf{a}$ we report the temporal frequencies, $\omega$, overlapped for multiple calibrations, as a function of their weight $a$. For each frequency, the weighting coefficient, $a$, is normalised with the largest one from the calibration. We perform the analysis with  $75$ (red), $90$ (green), $100$ (blue), $115$ (yellow) and $125$ (magenta) windows. The accuracy of the reconstruction depends on two thresholds, one in space and one in time, which we set to the same value, $\varepsilon$. The number of modes retained for the reconstruction is a direct consequence of the value of $\varepsilon$ and the number of windows.  The grey bars indicate the frequency dispersion of each group. In panel~$\mathbf{b}$ we illustrate the calibration we chose for this case, yielding a robust set of modes, shown with the same color coding of FIG.~$3$ of the main text.
 }
\label{fig:calibration}
\end{figure}

We provide a simplified description of our complex fluid flow identifying coherent structures that are organised in space and persistent in time. 
Since their robust detection is complicated by the broad range of space and time scales exciting the flow, we adopt the higher order dynamic mode decomposition (HODMD) \cite{vega-leclainche-2020}.
HODMD is a purely data-driven algorithm that recognises the most dynamically relevant large-scale motions in the flow. It extends the applicability of the classic dynamic mode decomposition (DMD) \cite{schmid-2010} to temporal data described by an almost incommensurable number of frequencies, implementing a delay embedding procedure. The algorithm analyses a sequence of observables (or windows) that overlapped constitute the original data. The number of windows is the most critical parameter to choose in our data-analysis, and it can be set in a wide interval proportional to the total number of observables. HODMD also performs a dimensionality reduction in both space and time. The low-dimensional space can be tuned manually, although the compression can deteriorate the reconstruction of the data. Here, we make sure that HODMD identifies the dominant structures in the flow.
FIG.~\ref{fig:calibration} exemplifies the calibration procedure for the case at $Re = 50$. We guarantee the robustness of our results by looking at groups of modes with similar frequencies (up to 20\% deviation) across several calibrations (75\% of the total). We find the same dominant frequencies regardless of the parameter choice, demonstrating the robustness of our analysis. The same calibration procedure provides similar results for the cases at $Re = 0.5$, $5$, $500$ and $1500$, where we fulfil a description of the complex flow dynamics with a minimum set of physically relevant modes.

\pagebreak

\begin{figure}
\centering
\includegraphics[scale=0.54]{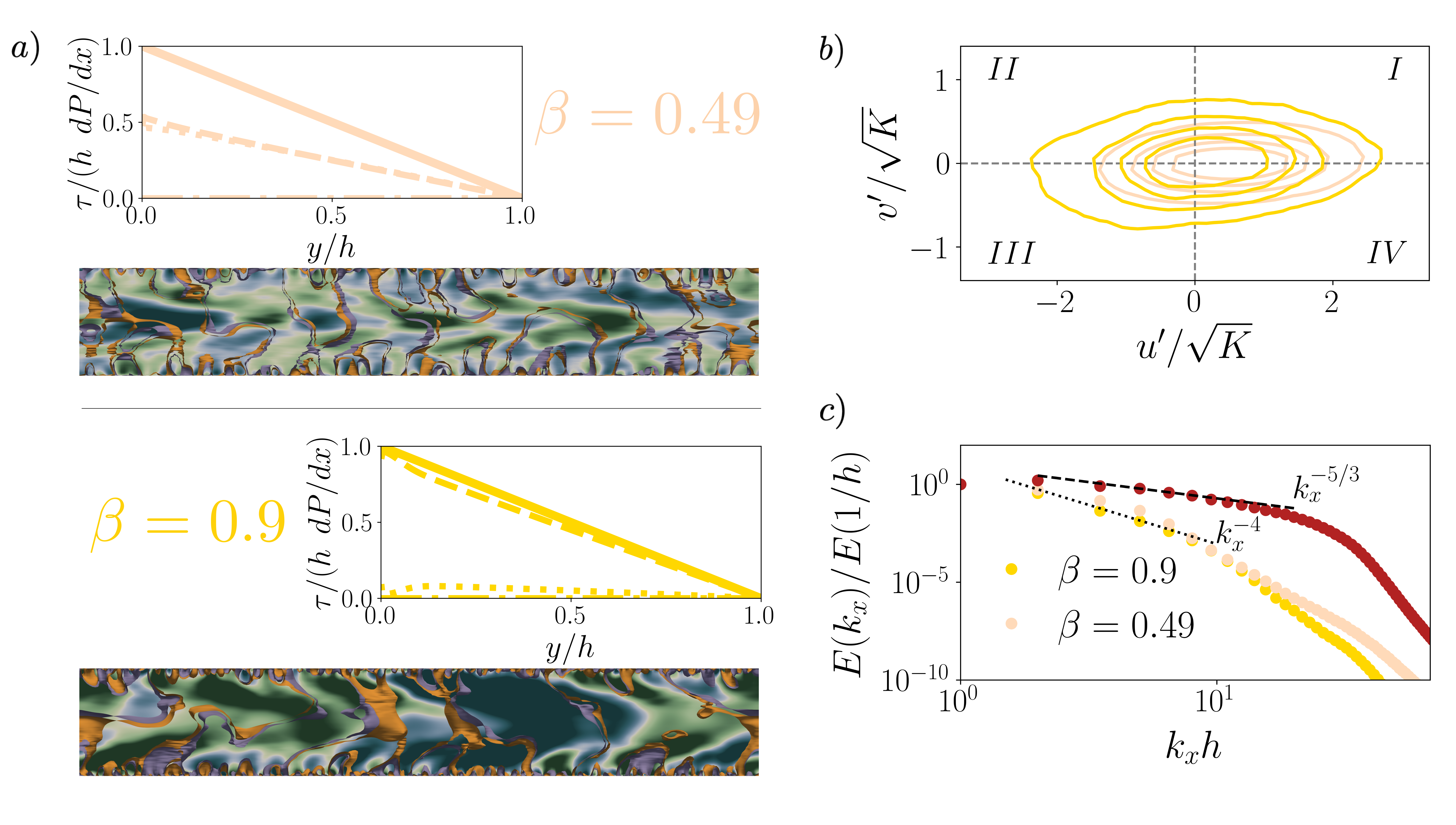}
\caption{ \textbf{Effect of the polymer concentration.} We characterize the flow fields at $Re=50$ for two different values of $\beta$. Panel~$\mathbf{a}$ shows side views of the channel, with the flow oriented from left to right, and the total shear stress balances along the wall-normal direction $y$. The views report the spanwise velocity fluctuations normalised with the mean turbulent kinetic energy, $w'/\sqrt{K}$, ranging from -0.3 (blue) to 0.3 (green); superimposed are the local isosurfaces of the wall-normal fluctuations, $v'/\sqrt{K}$, at $-0.05$  (orange) and $+0.05$ (violet). The total shear stress balances report the viscous (dashed), turbulent (dashed-dotted) and polymer (dotted) contributions, which sum up to the total shear stress $\tau$, exhibiting a linear behaviour (solid line). Increasing the value of $\beta$, the velocity fluctuations originated close to the wall occupy an increasingly thinner region, giving way to travelling waves in the center of the channel. At the same time, the polymer contribution to the total shear stress is depleted and the viscous one becomes dominant.
In panel~$\mathbf{b}$ we show the isolines of the J-PDFs associated to the normalised streamwise and wall-normal velocity fluctuations at $y=h/2$, choosing $\{0.1,0.3,0.5,0.7\}$ times the maximum  and adopting a color scale consistent with the plots in the other panels.
 The fluctuations appear decorrelated and dominated by the streamwise component at both values of $\beta$.
 The two different cases are also united by the streamwise spectra of the turbulent kinetic energy at the centerline, in panel~$\mathbf{c}$, normalised with their value at the beginning of the scaling region, $E(1/h)$. 
Both of them exhibit approximatively a $-4$ slope within their scaling region, opposed to the less steep trend of the TDR case \textcolor{black}{from FIG. 2e in the Letter (reported here with red dots)}, thus corroborating the unified view of EIT and ET.}
\label{fig:beta}
\end{figure}

\clearpage

\section{The effect of the polymer concentration}
We extend the body of simulations introduced in our Letter with two additional cases, thus spanning $\beta$ in $\{0.49,0.9,0.98\}$ at a fixed value of $Re=50$ and $De=50$.

\begin{figure}[h!]
	\centering
	\includegraphics[width=.9\textwidth]{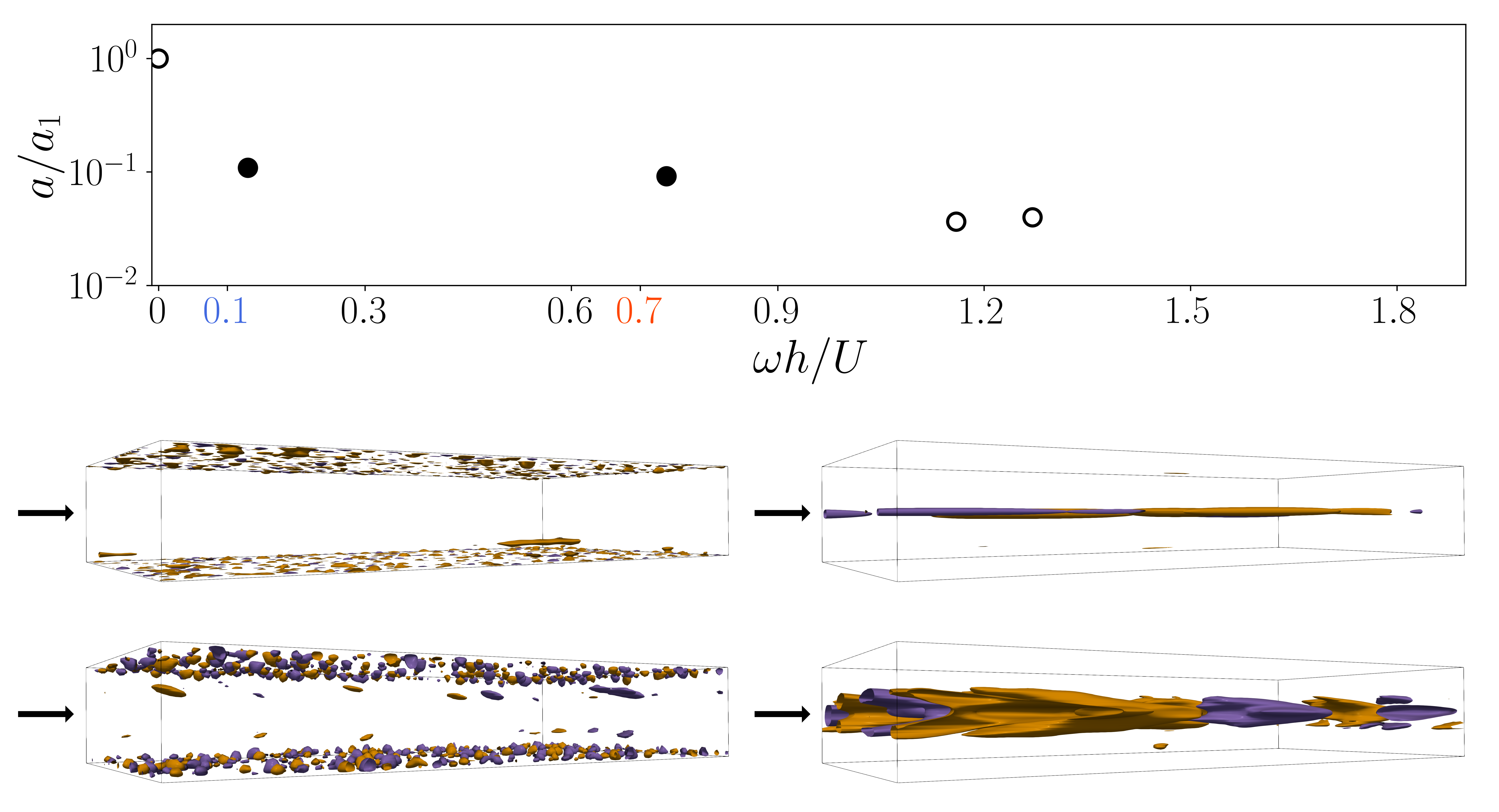}
	\put(-450,235){a)}
	\put(-450,115){b)}
	\put(-225,115){c)}
	\put(-355,120){$\omega \simeq 0.13$}
	\put(-130,120){$\omega \simeq 0.74$}
	\put(-260,65){\large \textit{u}}
	\put(-35,65){\large \textit{u}}
	\put(-260,4){\large \textit{w}}
	\put(-35,4){\large \textit{w}}
	\caption{\textbf{HODMD analysis for the case at $\mathbf{De = 50}$, $\mathbf{Re = 50}$ and $\mathbf{\boldsymbol{\beta} = 0.49}$.} Panel~$\mathbf{a}$ shows the robust frequencies, i.e., those recurring throughout several calibrations of the algorithm. We identify two frequencies, in filled black dots, close to those associated with the wall and center modes, respectively denoted in blue and red. We also report the streamwise (\text{u}) and spanwise (\textit{w}) velocity components of the two modes, in panel~$\mathbf{b/c}$: positive isosurfaces are colored in purple and negative isosurfaces in orange. The shape of the modes is affected by the low value of $\beta$ \cite{dubief-etal-2022,couchman-etal-2024}, although there is no significant difference in terms of frequency with the results reported for $\beta = 0.9$ in FIG.~3 of the Letter.}
	\label{fig:dmd}
\end{figure}

At $\beta=0.49$ ET occurs, and we thus observe the variation of the flow field and of the shear stress balance compared to the case at $\beta=0.9$ in FIG.~\ref{fig:beta}$a$.
For the lowest value of $\beta$, the flow is dominated by velocity fluctuations protruding from the wall region towards the center of the channel, justifying the steep increase of the polymer shear stress close to the wall.
The viscous shear is comparable to the polymer one throughout the channel, consistently with its high concentration.
Increasing $\beta$ to $0.9$, the polymer shear is globally depleted, but retains a higher gradient close to the wall as the velocity fluctuations in that region are still relevant. 
Despite the qualitative differences between the two cases,  the J-PDFs  of the streamwise and wall-normal velocity fluctuations at $y=h/2$ (in FIG.~\ref{fig:beta}$b$) and the scalings of the spectra of the turbulent kinetic energy at the centerline of the channel (in FIG.~\ref{fig:beta}$c$) remain qualitatively unchanged, confirming the generality of the turbulent state attained. 
HODMD modes with frequencies similar to those observed at $\beta=0.9$ are also found at $\beta=0.49$, as reported in FIG.~\ref{fig:dmd}: as in the case at $\beta = 0.9$, low frequency modes populate the region close to the wall, while high frequency ones are located in the center of the channel.
The HODMD results thus delineate a similar turbulent state at significantly different values of $\beta$, which mainly affects only the shape of the modes \cite{dubief-etal-2022,couchman-etal-2024}.

At $\beta=0.98$, instead, a laminar state is approached. 
This outcome is consistent with the predictions of stability analysis: at $\beta=0.98$ the mode located in the center of the channel is linearly stable \cite{buza-page-kerswell-2022} and non-linearly unstable \cite{buza-etal-2022}, while the mode at the wall is is stable \cite{beneitez-page-kerswell-2023} and disappears. 
The non-linear interaction between the center and wall modes \cite{beneitez-etal-2024} is thus lost and the center mode is no longer sustained, leading to the decay of the ET state in favour of the laminar one.

\pagebreak

\color{black}
\section{Variables and symbols}

Here we tabulate all the variables and few other symbols, following their order of appearance in the Letter. \\

\centering
\begin{tabular}{||c c||} 
 \hline
 \textbf{Symbol} & \textbf{Meaning} \\ [0.5ex] 
 \hline\hline
 $Re$ & Reynolds number \\ 
 \hline
 $De$ & Deborah number \\
  \hline
 $h$ & Half-channel height \\
 \hline
 $x$ & Streamwise coordinate \\
 \hline
 $y$ & Wall-normal coordinate \\
 \hline
 $z$ & Spanwise coordinate \\
 \hline
 $\mathbf{u}$ & Fluid velocity \\
 \hline
 $\rho$ & Fluid density \\
 \hline
 $p$ & Hydrostatic pressure \\ 
 \hline
 $\mu_s$ & Dynamic viscosity of the Newtonian solvent \\
 \hline
 $\mathbf{f}$ & Forcing \\
 \hline
 $\mathbf{T}$ & Non-Newtonian stress tensor \\
 \hline
 $\mathbf{C}$ & Polymer conformation tensor \\
 \hline
 $L_{max}$ & Maximum extensibility of the polymer chains \\
 \hline
 $\mathbf{I}$ & Tensorial identity \\
 \hline
 $\mu_p$ & Dynamic viscosity of the polymer \\
 \hline 
 $\lambda$ & Relaxation time of the polymer \\ 
 \hline
 $\beta$ & Solvent-to-total viscosity ratio \\
 \hline
 $U$ & Mean velocity within the channel \\
 \hline
 $\nu$ & Total kinematic viscosity of the polymeric solution \\
 \hline
 $El$ & Elasticity number \\
 \hline
 $\tau$ & Total shear stress \\
 \hline
 $d_xP$ & Driving pressure gradient \\
 \hline
 $T_{xy}$ & Shearing component of the viscoelastic stresses \\
 \hline
 $\langle \cdot \rangle$ & Averaging in time and along the homogeneous directions \\
 \hline
$\cdot '$ & Fluctuations from the mean \\
\hline
$K$ & Mean turbulent kinetic energy in the channel \\
\hline
$E$ & Turbulent kinetic energy at the centreline of the channel \\
\hline
$k_x$ & Streamwise wave number \\
\hline
$\omega$ & Frequency \\ 
 \hline
\end{tabular}
\centering

\hspace{5cm}
\color{black}

\bibliography{./Wallturb.bib}